\documentstyle[twoside,fleqn,espcrc2,epsf]{article}

\title{Effect of Order-Parameter Suppression on Scattering by Isolated
Impurities in Asymmetric Bands}

\author{W. A. Atkinson$^{\rm a}$,
P. J. Hirschfeld\address{Department of Physics, University of
Florida, PO Box 118440, Gainesville FL 32611} and
A. H. MacDonald\address{Department of Physics, Indiana University,
Bloomington IN 47405}}

\begin{document}

\begin{abstract}
The single-impurity problem in $d$-wave superconductors with
asymmetric bands is discussed.  The effect of local order parameter
suppression near the impurity is to shift the 
quasiparticle resonance.  Contrary to previous work
[A.\ Shnirman {\em et al.}, Phys. Rev. B {\bf 60}, 7517 (1999)]
we find that the direction of the shift is not universally towards
the strong scattering limit.
\end{abstract}

\maketitle

There have been many theoretical studies of the local density of states (DOS)
near an isolated impurity in a $d$-wave superconductor recently.
The main conclusions of this body of work are that
the $d$-wave symmetry of the order parameter (OP) is reflected in the
spatial structure of the DOS\cite{Scalapino}, and that a pair 
of quasiparticle resonaces at energies $\pm\Omega_0$ occur within the
$d$-wave gap\cite{Balatsky}.  While these conclusions appear to be 
robust, the detailed structure of the DOS and the magnitude of $\Omega_0$
depend sensitively on details of the band structure\cite{Joynt,atkinson} and 
impurity potential.

In this work, we show that local suppression of the OP
near the impurity acts as an important anomalous scattering potential.
With the exception of Refs.~\cite{Hettler,Shnirman}, this effect has generally
been ignored.  In \cite{Shnirman}, a relatively simple and physically
appealing model of OP suppression was used to study the
resonance structure of the 1-impurity scattering T-matrix. It was found that
scattering from inhomogeneities in the OP renormalizes
the energies $\Omega_0$ towards the unitary limit $\Omega_0 = 0$ while
leaving qualitative aspects of the resonances unchanged.  The authors
suggested this as an explanation for the apparent proximity of Zn to
the unitary limit in high $T_c$ materials.  Here, we extend this
discussion to the case of asymmetric bands, which have been
shown previously\cite{Joynt} to be important in the single impurity
problem.  Our main conclusion is that the renormalisation of $\Omega_0$
is not generally towards the unitary limit, except in the unphysical
case of perfectly symmetric bands.

The calculations are based on an exact T-matrix method, described
elsewhere\cite{Byers}.  The model consists of a tight-binding
lattice with nearest neighbour hopping, nearest neighbour
pairing, and a single, point-like-impurity at the origin.
The Hamiltonian is:
\begin{eqnarray}
{\cal H} &=& -t \sum_{\langle i,j \rangle} c^\dagger_{i \sigma} c_{j\sigma}+
 \sum_i (u_0 \delta_{i,0} - \mu) n_i \nonumber \\
&&+ \sum_{\langle i,j \rangle} \Delta_{ij} c^\dagger_{i\uparrow}
c^\dagger_{j\downarrow} + \Delta_{ij}^\dagger c_{i\downarrow}
c_{j\uparrow}
\end{eqnarray}
where $c_{i\sigma}$ is the annihilation operator for an electron on
site $i$ with spin $\sigma$ and $n_i$ is the electron density at site
$i$.  The hopping matrix element $t$ is the unit of energy throughout
this work.  Self-consistent solutions for $\Delta_{ij}$  from
the Bogoliubov-deGennes equations show that $\Delta_{ij}$ is
suppressed along bonds connected to the impurity site, and regains
its homogeneous clean-limit value within a few lattice constants.  The
inhomogeneous portion of $\Delta_{ij}$ is extracted, and treated as a
spatially-extended anomalous scattering potential, additional to the
on-site impurity potential. The T-matrix, which relates the scattering
states to the eigenstates of the impurity-free system, is found by
solving the Lippmann-Schwinger equation.

\begin{figure}[t]
\begin{center}
\epsfxsize \columnwidth
\epsffile{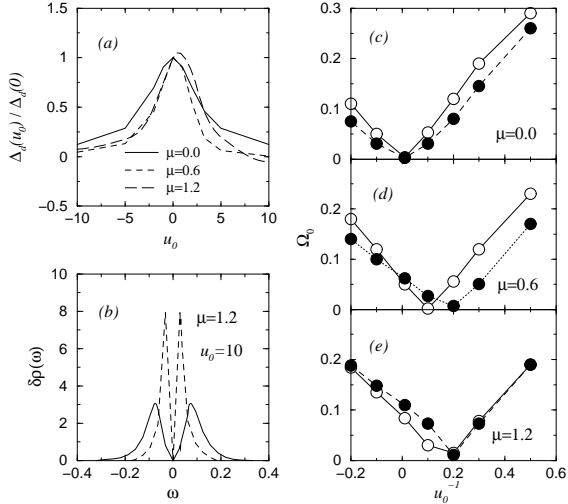}
\caption{Effect of OP suppression on the DOS. (a)
$d$-wave OP at the impurity site relative to impurity-free value.
(b) change in the DOS produced by scattering
with (solid) and without (dashed) OP suppression.
(c)--(e) Resonance peak position vs.\ scattering potential. 
Closed (Open) symbols indicate 
(no) OP suppression.}
\label{fig1}
\end{center}
\end{figure}

In Fig.~\ref{fig1}, the effect of band asymmetry on the spatially-%
integrated DOS is
shown.  The controlling parameter is $\mu$, with half-filling
($\mu=0$) corresponding to a symmetric band.  From (a), we see that
the OP suppression is a weakly asymmetric function of
$u_0$ for $\mu\neq 0$.  Typical quasiparticle resonances are
illustrated in (b).  For the model
parameters chosen, the $d$-wave OP is $\Delta_d = 0.39$
and the gap edge in the DOS is $\approx 0.5$.  The spectral weight
within the resonances at $\pm \Omega_0$ is transferred primarily from the gap
edges (not shown)\cite{Byers}.  The broadening of the resonances is determined
by the background density of states $\rho_0(\Omega_0)$ and specifies
the lifetime of a quasiparticle near the impurity before it leaks
away through continuum states.  The two resonances shown in (b)
occur for the same set of model parameters, with and without 
inclusion of the scattering from the inhomogeneous OP field, and
are an instance in which OP suppression drives the resonance {\em away}
from $\Omega_0=0$.

We plot the positions of the 
resonance peaks in Figs.~(c)-(e), for different values of $\mu$ and
$u_0$.  In (c), the band is symmetric, and the shift of $\Omega_0$
is always towards the strong scattering limit, as predicted
by \cite{Shnirman}.  For $\mu = -0.6$, the bands become asymmetric
and two qualitative aspects are changed:  the unitary ($\Omega_0 = 0$)
and strong impurity ($u_0^{-1}=0$) limits no longer coincide,
and the direction of the shift in $\Omega_0$ is nonuniversal.  
The former effect has been discussed at length in \cite{Joynt},
but the latter effect is new. For
bands which are still more asymmetric (eg.\ $\mu=-1.2$) an
additonal factor becomes important;  scatterers which are near the
unitary limit may actually have sufficiently small values of 
$u_0$ that the amount of OP suppression is small.  Hence,
the shift in $\Omega_0$ for $u_0^{-1}>0.2$ is quite small.

We have shown that the simple ansatz for nearest-neighbour 
OP suppression made by \cite{Shnirman} gives a good
approximation to both the impurity resonance position and
the momentum-space structure of the inhomogeneous OP in a $d$-wave
superconductor with symmetric bands.  Qualitatively different results
were obtained for more realistic asymmetric bands.  We will
report on novel effects of OP suppression in bulk disordered systems
elsewhere\cite{atkinsonI}.


\begin{thebibliography}{Fehrnenbacher}
\bibitem{Scalapino} J. M. Byers, M. E. Flatt\'e, and D. J. 
Scalapino, Phys. Rev. Lett. {\bf 71} 3363 (1993).

\bibitem{Balatsky} A. V. Balatsky, M. I. Salkola, and A. Rosengren,
Phys. Rev. B {\bf 51} 15 547 (1995); A. V. Balatsky and
M. I. Salkola, Phys. Rev. Lett. {\bf 76}, 2386 (1996).

\bibitem{Joynt} R. Joynt, J. Low Temp. Phys. {\bf 109}, 811 (1997).

\bibitem{atkinson} W. A. Atkinson and A. H. MacDonald, 
Science {\bf 285}, 57 (1999).

\bibitem{Hettler} M. H. Hettler and P. J. Hirschfeld,
Phys. Rev. B {\bf 59}, 9606 (1999).

\bibitem{Shnirman} Alexander Shnirman, \.{I}nanc Adagideli, Paul
M. Goldbart, and Ali Yazdani, Phys. Rev. B {\bf 60}, 7517 (1999).

\bibitem{Byers} M. E. Flatt\'e and J. M. Byers, Solid State Phys. {\bf
52}, 137 (1999).  

\bibitem{atkinsonI} W. A. Atkinson, P. J. Hirschfeld, and
A. H. MacDonald (unpublished).






\end{thebibliography}
\end{document}